\documentclass[traditabstract]{aa} 
\usepackage{graphicx}
\usepackage{txfonts}
\begin{document}
   \title{U Sco 2010 outburst: observational evidence of an underlying ONeMg white dwarf\thanks{Based on observations carried out at the ESO's VLT+X-Shooter under program 284.D-5041.}}

   \subtitle{}

   \author{E. Mason
          \inst{1}
          }

   \institute{Space Telescope Science Institute (STScI),
              3700 S. Martin Dr., Baltimore, MD 21218;
              \email{emason@stsci.edu}
              }

   \date{Received June 8, 2011; accepted Jul 18, 2011}

 
  \abstract
{This paper presents U\,Sco nebular spectra collected in the period March-May 2010 after the binary outburst on Jan 28, 2010. The spectra display strong [Ne{\sc v}] and [Ne{\sc iii}] lines that can be used to compute the relative abundance of [Ne/O]. The value obtained ([Ne/O]=1.69) is higher than the typical [Ne/O] abundance found in classical novae from CO progenitors and suggests that U\,Sco has a ONeMg white-dwarf progenitor. It follows that U\,Sco will not explode as a SN\,Ia but rather collapse to become a neutron star or a millisecond pulsar.} 

   \keywords{stars: individual: U\,Sco --novae, cataclysmic variables  }

   \maketitle
%

\section{Introduction}

The outburst of U\,Scorpii (U\,Sco) on Jan 28, 2010 at $V_{\sc max}=8.05$\,mag (Munari et al. 2010) is the tenth outburst recorded for this nova and, thanks to the Schaefer monitoring and alert (see e.g. Schaefer et al.\ 2010), the best-studied of this kind together with the outburst of 1999. 
The importance of observing U\,Sco and recurrent novae (RNe) outbursts in general resides in the opportunity to understand the kinematics and the composition of RNe ejecta in relation to both classical novae (CNe) and supernovae (SNe). While observations (Della Valle and Livio 1996) show that RNe are not the main contributors to the class of SN\,Ia progenitors, numerous theoretical works consider  RNe within the single degenerate scenario as valid progenitors of type Ia SNe (e.g. Hachisu et al.\ 2000; Hachisu and Kato 2002; Livio 2000 and reference therein; Nomoto et al.\ 2000; see also Podsiadlowski 2008). 
To better understand this problem, I collected X-Shooter multi-wavelength medium resolution spectra of U\,Sco during its late decline from the plateau phase down to quiescence. The goal was to characterize the physical parameters of the ejecta when they are optically thin and compute  abundances, which are a discriminating factor between eroded white dwarfs (WDs) vs. non-eroded WDs.  In this letter, I present nebular spectra of U\,Sco taken 45, 73, and 104 days after the outburst and discuss the measured [Ne/O] abundance value in the attempt to identify the underlying WD. A more detailed discussion of the U\,Sco 2010 outburst spectral evolution will be presented in a separate paper (Mason et al. in preparation), which collects a larger sample of spectra.


\section{Observation and data reduction}

X-Shooter is the first of the second-generation instruments installed on the VLT (D'Odorico et al.\ 2006). It is a three-arm spectrograph that allows medium-resolution spectroscopy in the wavelength range 300-2500 nm in a single exposure. 
The observations were performed in service mode within ideal time windows for each epoch. The spectra were taken on March 15 (i.e. +46 days from maximum, at orbital phase 0.07), April 11 (+73 days, at orbital phase 0.01), and May 12 2010 (+104 days, at orbital phase 0.16). The instrument setup on March 15 was slit 0.8$^{\prime\prime}$, 0.7$^{\prime\prime}$ and 0.6$^{\prime\prime}$ for the UVB, VIS, and NIR arms, respectively;  it was slit 1.0$^{\prime\prime}$, 0.9$^{\prime\prime}$ and 0.6$^{\prime\prime}$ for the April and May observations. The CCD readout was always 100\,kHz (high gain), but 1$\times$2 binning was adopted for the April and May observations. The observing strategy involved nodding along the slit with a nod-throw of 5$^{\prime\prime}$. The slit was always oriented along the parallactic angle at the beginning of each exposure and the selected telluric standard star (also observed nodding along the slit) was the Hipparcos star Hip 82254, a B3\,{\sc v} star of magnitudes {\it B}=6.76 {\it V}=6.81 and {\it H}=6.92. The data were reduced using the instrument pipeline, version 1.2.2, in ``physical model mode'' (Modigliani et al.\ 2010; see also Bristow et al.\ 2010) for image pre-processing, sky-subtraction and spectral rectification. Order extraction and merging, telluric correction, extinction correction, and flux calibration were performed with IRAF. 
The spectrophotometric standard EG~274, observed on May 12, was used to calibrate all spectra (see Vernet et al., 2008, for a UV to NIR database of spectrophotometric standard stars). The flux-calibrated spectra were aligned to the VIS arm and scaled to U\,Sco J magnitude at the corresponding epoch (Schaefer, private communication). The spectra were then dereddened adopting {\it E(B$-$V)}=0.15 (from the equivalent width of the Na D$_1$ interstellar absorption, Munari and Zwitter 1997) and the upper and lower limits of 0.25 and 0.00 mag, respectively, which account for the large uncertainties resulting from the combination of different methods (e.g. Diaz et al. 2010).  

\section{The nebular phase and the forbidden emission lines}
   \begin{figure}
   \centering
   \includegraphics[width=7cm,angle=270]{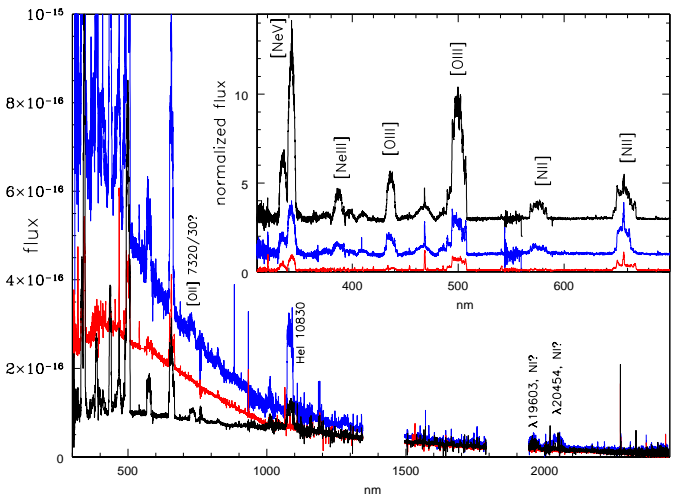}
   \caption{U Sco X-Shooter spectra: Mar 15 in blue, Apr 11 in black and May 12 in red. In the inset: a portion of the same spectra (UVB+VIS) showing the nebular forbidden transitions. The spectra in the inset have been normalized to unity and offset vertically for clarity. They are sorted in order of their nebular line intensity for clarity. The displayed spectra have been median-filtered.}
              \label{spc}%
    \end{figure}
%


The sequence of collected spectra (Fig.\,1) covers the U\,Sco outburst light curve from the second plateau after the super soft phase has ended ($V\geq$16.5\,mag) to the quiescent phase ($V\geq$18\,mag; see, for example, Munari et al. 2010 and Diaz et al. 2010 for the U\,Sco 2010 outburst light curve). The spectra show that beginning Mar\,15, U\,Sco entered the  nebular phase, displaying strong and broad emission lines of [N{\sc ii}], [O{\sc iii}], [Ne{\sc iii}] and [Ne{\sc v}]. The nebular phase of U\,Sco was reported already by Diaz et al.\ (2010), who observed [N{\sc ii}], [O{\sc iii}], and [Ne{\sc iii}] starting from day +75 after maximum. Though it has been claimed that U\,Sco does not develop forbidden transitions (e.g. Warner 1995)
and that the 2010 outburst was the first time that the nova displayed a nebular spectrum (e.g. Diaz et al. 2010), neither statement is accurate. First, because the observations performed during the past outbursts were not sufficiently extended in time after the maximum; second, because the average spectrum of Thoroughgood et al.\ (2001, their figure 1) clearly shows a broad composite emission, which should be identified with [O{\sc iii}] $\lambda$5007. Its profile is very similar to that reported in this paper (see Fig.1 inset).

The forbidden transitions appear when the U\,Sco continuum has significantly dimmed and the nova is about 3 magnitudes fainter than during the first plateau phase. At this epoch the continuum varies not only with the time since maximum, but also with the orbital phase (see Fig.\,1). The April spectrum  was taken close to the eclipse time (at orbital phase 0.01) and shows a flat continuum. The March and May  spectra (centered at orbital phase 0.07 and 0.16, respectively) are characterized by a blue continuum. This can be explained by the hot blue component (the white dwarf itself or an accretion hot-spot) being masked by the secondary star at the time of the April observation.

The evolution of the broad emission lines from the ejecta is independent of the orbital phase
, but their intensity is maximum when the continuum strength decreases. The three nebular spectra also show that the resonant transitions constantly weaken in time relative to the forbidden transitions, until they almost completely disappear by May\,12. This is indicative of a progressively lower density in the ejecta, though this is always relatively high when compared to the densities in planetary nebulae. 
The analysis of the line profiles shows that the H$\alpha$ emission dominates the ``6563 blend'' in the March and April spectra, and that it is significantly weaker than the [N{\sc ii}]$\lambda$6584 emission only in the May spectrum. Hence, only in the latter spectrum it is possible to use the [N{\sc ii}] lines ratio to constrain the ejecta density and temperature. At this epoch the [N{\sc ii}]\,$\lambda$(6584+6548)/$\lambda$5755 flux ratio is $\sim$6, implying that collisions are contributing to the line formation and that the gas densities are higher than $10^5$\,cm$^{-3}$ (see e.g. Osterbrock and Ferland 2006). 
High gas densities are suggested also by the [O{\sc iii}]\,$\lambda$(5007+4959)$\lambda$4363  flux ratio\footnote{ In computing these ratios, the conservative H$\gamma$ flux F$_{H\gamma}=0.85\times$F$_{H\beta}$ has been subtracted from the [O{\sc iii}] $\lambda$4363 emission, though the $\lambda$4363 profile does not show strong evidence of a blend. The H$\gamma$/H$\beta$ ratio has been inferred from flat Balmer decrement H$\delta$/H$\beta\geq$0.7 measured in both the March and April spectra. The derived flux ratios should, therefore, be taken as an upper limit. No H$\gamma$ fractional contribution to the 4363 line has been assumed in the May spectrum, because, at that time, no broad emission component from the ejecta is detectable in any of the Balmer lines (and in H$\delta$, in particular).}, which is  2.97, 4.79 and 9.37 in the March, April and May spectra, respectively, and indicates densities  $\geq5\times10^{6}$-10$^7$\,cm$^{-3}$ for temperatures in the nominal range 14000-10000\,K. Figure 2 plots the diagnostic diagram for the [O{\sc iii}] and [N{\sc ii}] flux ratios. Note that the lower temperatures and densities indicated by the [N{\sc ii}]$\lambda\lambda$6584,6548/[N{\sc ii}]$\lambda$5755 flux ratio are consistent with the fact that the [N{\sc ii}] transitions typically form in the outer and cooler shell of the expanding ejecta (e.g. Osterbrock and Ferland 2006).

   \begin{figure}
   \centering
   \includegraphics[width=7cm]{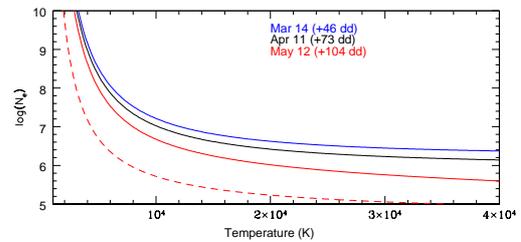}
      \caption{Diagnostic diagram for the flux ratios (F$_{4959}+$F$_{5007}$)/F$_{4363}$ (solid lines) and (F$_{6548}+$F$_{6583}$)/F$_{5755}$ (dashed lines) at the three epochs Mar 15 (blue), April 11 (black), and May 12 (red). Each line represents the loci of density and temperature that match the observed flux ratios. Fluxes correspond to the dereddened spectra. 
              }
         \label{diag}
   \end{figure}

Though accurate elemental abundance determination requires the combined modeling of UV and optical observations  (e.g., Schwarz 2002), a first oder approximation of the relative abundances can be computed by using the flux ratios of emission lines from ions with similar ionization potential energies, similar critical densities, and the same excitation mechanism (Kingdon and Williams 1997). This method has the advantage of being fairly insensitive to temperature uncertainties (temperatures that differ by a factor of 2 imply uncertainties of $\sim$20\% in the abundance) and has been tested against models for a range of temperatures and densities. In the case of the U\,Sco nebular spectra, it is possible to compute the [Ne/O] abundance from the flux ratio of the lines [Ne{\sc iii}]$\lambda$3869 and [O{\sc iii}]$\lambda$5007,
in both the March and April spectra. In the high-density limit (e.g. Dopita 2001) the line flux can be written as 
\begin{equation}
F_{ij}=N_iE_{ij}A_{ji}{{g_j}\over{g_i}}exp[{{-E_{ij}}\over{kT}}] ,
\end{equation}
where $i$ and $j$ are the two levels of the transition, $N_i$ is the density of level $i$, $E_{ij}$ is the energy of the transition, $A_{ij}$ is the transition probability, $g_i$ and $g_j$ are the statistical weights of the states, $k$ is the Boltzmann constant, and $T$ the gas temperature. Adopting T$_e\sim$12000\,K, as derived from the above considerations and Fig.\,2, one obtains the relative abundances of [Ne/O]=1.97 (March) and [Ne/O]=1.69 (April), as reported in Table\,1. Kingdom and Williams (1997) established that the [Ne{\sc iii}]$\lambda$3869/[O{\sc iii}]$\lambda$5007 pair is a good indicator of the [Ne/O] abundance for gas densities $<10^6$\,cm$^{-3}$, while U\,Sco ejecta have densities that are about one order of magnitude higher. To quantify the uncertainty associated to the method in this case, I applied it to V382\,Vel (Della Valle et al. 2002, their table\,3), a very fast nova, which developed strong [Ne{\sc iii}] optical emission lines in a high-density ejecta ($\simeq 10^7$\,cm$^{-3}$), similarly to U\,Sco. Comparing V382\,Vel [Ne/O] abundance derived from the $\lambda$3869/$\lambda$5007 flux ratio, with that obtained via photo-ionization modeling (Shore et al. 2003), one obtains relative errors in the range 25-50\%, depending on the epoch. Hence, the U\,Sco [Ne/O] abundances could be as low as 0.99$^{+0.01}_{-0.10}$ and 0.85$^{+0.01}_{-0.06}$ for the March and April spectra, respectively. This would not change the conclusion derived in the next section.
 
\begin{table}
\scriptsize
 \centering
  \caption{Integrated emission line fluxes of the ejecta as measured in the X-Shooter dereddened spectra. The fluxes measured on the spectra before correction for reddeining are reported in  brackets.  The [O{\sc iii}]$\lambda$5007 flux has been computed assuming the theoretical transition probability of 3 after subtraction of the H$\beta$ contribution.  Fluxes in the table have errors in the range 1-30\%, depending on the line strength; while the lower and upper limit in the [Ne/O] abundances have been computed measuring the spectra corrected for reddening assuming {\it E(B-V)}=0.25\,mag and the uncorrected spectra, respectively, also adding the appropriate uncertainty on the measured line flux. }
  \begin{tabular}{lccc}
  \hline
   line ID \& $\lambda$ (\AA)  & \multicolumn{3}{c}{Flux ($\times 10^{-15}$ erg\,cm$^{-2}$\,s$^{-1}$\,\AA$^{-1}$) } \\
  &  March 15 &  April 11 & May 12  \\
 \hline
$[$Ne{\sc v}$]$(1) 3343 & 42.5 (20.8) & 21.2 (10.3) & 5.8 (2.8) \\
$[$Ne{\sc v}$]$(1) 3426 & 119.0 (60.2) & 56.0 (27.9) & 15.4 (7.5) \\
$[$Ne{\sc iii}$]$(1) 3869 & 41.2 (18.7) & 12.4 (6.1) & -- \\
$[$Ne{\sc iii}$]$(1) 3967 & 13.8 (5.7) & 3.7 (1.8) & -- \\
H$\delta$ & 12.0 (5.6) & 2.7 (1.1) & -- \\
$[$O{\sc iii}$]$(2) 4363 & 60.2 (33.2) & 19.4 (11.4) & 2.6 (1.4) \\
$[$O{\sc iii}$]$(1) 5007+4956  & 151.0 (84.0) & 81.4 (49.3) & 22.8 (14.1)  \\
$[$O{\sc iii}$]$(1) 5007 & 102.0 (63.0) & 59.8 (37.0) & 16.9 (10.5) \\
H$\beta$ 4861 & 17.0 (11.3) & 3.5 (2.5) & 0.6 (0.4) \\
$[$N{\sc ii}$]$(3) 5755 & 18.2 (12.2) & 9.8 (6.7) & 2.0 (1.6)\\
H$\alpha$+$[$N{\sc ii}$]$(1) & 58.3 (41.9) & 22.2 (15.8) & 12.0 (8.7) \\
\hline
$[$Ne/O$]$ & 1.97$^{+0.03}_{-0.19}$ & 1.69$^{+0.02}_{-0.12}$ & - \\
\hline
\end{tabular}
\end{table}

\section{Discussion and conclusion: the ultimate U Sco fate}

The general consensus about SN\,Ia progenitors is that they originate from CO WDs accreting matter up to the Chandrasekhar limit either via stellar merging (double degenerate scenario where two CO WDs coalesce) or via mass transfer from a less evolved companion (single degenerate scenario). Within the single degenerate scenario the accreting WD ought to be massive (i.e. close to the Chandrasekhar limit), and accreting at a high rate ( $\geq$10$^{-8}$-10$^{-7}$\,M$_\odot$\,yr$^{-1}$, Nomoto et al. 2007 and reference therein, see also Yaron et al. 2005). Hence, RNe whose frequent outbursts are explained by massive WDs and high mass-transfer rates (e.g. Truran et al. 1988) remain  viable candidates, contrary to what  is suggested by the population synthesis simulations (Di Stefano 2010 and reference therein) and their census (Della Valle and Livio 1996). U\,Sco, having a WD $>$1.37M$_\odot$ (Hachisu et al. 2000; Thoroughgood et al. 2001) and accreting at a rate $>10^{-7}$\,M$_\odot$\,yr$^{-1}$ (Hachisu et al. 2000; Matsumoto et al. 2003), seems very likely to explode as a SN\,Ia. However, it is also general consensus that an accreting ONeMg WD cannot lead to a SN\,Ia explosion but, eventually, to a core collapse and the formation of a neutron star or a millisecond pulsar (e.g. Nomoto and Kondo 1991). 
The high U\,Sco [Ne/O] abundance (when compared to the solar values of $-$0.76) is intriguing both because it points to dredged-up material and a possibly eroded WD, and because it provides information about the composition/nature of the underlying WD. 
Dredged-up material is typically associated with an eroded WD because, according to thermo-nuclear (TNR) computations (e.g. Prialnik and Livio 1995), anything that is mixed within the H-layer where the TNR ignites is ejected into the circumbinary space. If at any outburst the mass of the ejecta matches or exceeds the accreted mass, then the WD (independent of its composition) cannot increase in mass and reach the Chandrasekhar limit. On the other hand, if the primary star is an ONeMg WD, it will never explode as a SN\,Ia even in case of mass gain up to the Chandrasekhar limit of 1.4M$_\odot$. Hence, it has become critical to establish the composition of the accreting primary for U\,Sco because despite its WD mass, high accretion rate and small ejecta mass (a few $10^{-7}$\,M$_\odot$, Iijima 1999; Anupama and Dewangan 2000, but see also Diaz et al. 2010), it might not explode as a SN\,Ia but undergo core collapse at the very most. 

\begin{table}
\scriptsize
 \centering
  \caption{Ne abundances for a sample of classical novae. The [Ne/O] abundances have been computed assuming the solar values in Asplund et al. (2009). }
  \begin{tabular}{lccc}
  \hline
nova   & WD & [Ne/O] & ref. \\
\hline
V693 CrA  & ONeMg & +1.21 &  Vanlandingham et al. (1999) \\
V1370 Aql  &  ONeMg  &  +1.76 & Livio \& Truran (1994) \\
V1974 Cyg  & ONeMg &  +0.47 & Vanlandingham et al. (2005) \\
V382 Vel   & ONeMg &  +0.66 & Shore et al. (2003) \\
V4160 Sgr  & ONeMg &  +0.69 & Schwarz et al. (2007a) \\
V838 Her  & ONeMg &  +1.61 & Schwarz et al. (2007a) \\
QU Vul    & ONeMg &  +0.92 &  Schwarz (2002)\\
V1065 Cen   & ONeMg & +0.74 & Helton et al. (2010) \\
LMC 1990 N.1 & ONeMg  &  +0.45 & Vanlandingham et al. (1999) \\
U Sco  & ONeMg & +1.69 & this paper \\
RR Pic    & ? & +0.94  & Livio \& Truran (1994)\\
V977 Sco   & ? & +0.60 & Livio \& Truran (1994) \\
V1186 Sco   & CO & $-$0.42 & Schwarz et al. (2007b) \\
LMC 1991 & CO & $-$1.65 & Schwarz et al. (2001) \\
QV Vul  &  CO & $-$0.96 & Livio \& Truran (1994)\\
HR Del   & CO & $-$0.53 & Livio \& Truran (1994)\\
V1500 Cyg    & CO & $-$0.069 & Livio \& Truran (1994) \\
V1688 Cyg   &  CO  & $-$0.62 & Livio \& Truran (1994)\\
GQ Mus  & CO & $-$1.09 & Livio \& Truran (1994) \\
PW Vul  & CO &  $-$0.67 & Livio \& Truran (1994)\\
V842 Cen   & CO  & $-$0.86 & Livio \& Truran (1994) \\
V827 Her   &  CO  & $-$0.73 & Livio \& Truran (1994) \\
V2214 Oph   &  CO  & +0.11 & Livio \& Truran (1994) \\
V443 Sct    & CO  & $-$1.04 & Livio \& Truran (1994) \\
\hline
\end{tabular}
\end{table}

Livio and Truran (1994) cautioned observers about identifying ONeMg WD progenitors in CN ejecta that show Ne emission lines in their optical spectra. They showed that moderate Ne abundances (with respect to solar) can be explained either by abundance uncertainties or dredged-up material from the underlying CO WD, or with the breakout of the CNO cycle under special conditions. The authors identified a group of ``true ONeMg WDs'' in those novae that showed extreme enrichment of Ne and heavier elements. 
Table~2 lists the novae used by Livio and Truran (1994) for their analysis, as well as a number of CNe whose WD and abundances have been determined via photo-ionization modeling of UV and optical observations, simultaneously, by Schwarz and collaborators. 
The table reports the [Ne/O] abundances for each nova as well as the U\,Sco April\,11 relative abundance derived in this paper. From Table~2 it is evident that the CNe hosting a CO WD are characterized by [Ne/O] abundances $\leq$0, while those possessing an ONeMg WD have [Ne/O]$>$0. This is also shown in Fig.3, which plots the distribution of CNe as a function of their [Ne/O] abundance. The shaded histogram of Fig.3 corresponds to the ``fiducial sample'' of ONeMg CN white dwarfs; while the black areas correspond to two CNe that are considered as dubious by Livio and Truran (1994) on the basis of the high measured Ne abundances but relatively low values of the total heavy elements enrichment.  It should be noted that there were initially three dubious cases identified by Livio and Truran (1994), but IUE observations of Nova LMC\,1990\,N.1 (Starrfield et al.\ 1992, Vanlandingham et al. 1999) confirmed it to be an ONeMg nova similar to V463\,CrA. 

   \begin{figure}[t]
   \centering
   \includegraphics[width=7cm]{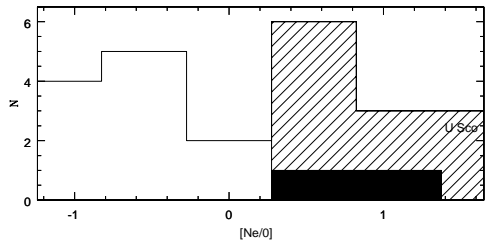}
      \caption{The histogram of the CNe distribution as a function of [Ne/O] abundance. The white area represents the CO novae, the shaded are represents the ONeMg novae, while the black area represents the two CNe which show extreme Ne enrichment but relatively small values for the total heavy elements enrichment. See text for more details.}
         \label{histo}
   \end{figure}

The value of [Ne/O]$>$1 determined in this paper for U\,Sco places the binary among the ``true Ne novae'', hosting an ONeMg WD.
It should also be noted that U\,Sco was observed by IUE during the 1979 outburst and on that occasion Williams et al. (1981) reported absorption lines and P-Cyg profiles from C{\sc iv}, Si{\sc iv} and N{\sc v} in their early epoch data. All ONeMg novae - contrary to the CO novae- show a P-Cyg profile phase in their UV spectra after the iron curtain phase and before the transition to the nebular spectrum (Shore 2008). The UV lines displayed at this stage are Si{\sc iv} 1400, C{\sc iv} 155 0, Al{\sc iii} 1860, and Mg{\sc ii} 2800 and show a saturated absorption trough with very high terminal velocity (Shore 2008). The IUE spectra of U\,Sco taken +4 and +6 days after the maximum clearly show P-Cyg profiles with broad absorption troughs that are very  similar to the IUE spectra of the ONeMg novae discussed by Shore (2008). Hence, U\,Sco should be regarded as a recurrent nova hosting a massive ONeMg WD (M$_{WD}\sim$1.37-1.55 M$\odot$, Hachisu et al.\ 2000 and Thoroughgood et al.\ 2001). The WD will undergo core collapse and will not explode as a SN\,Ia, unless the current models about accreting ONeMg WDs are significantly in error and the U\,Sco mass accretion rate and WD mass prove to be substantially smaller.

It is interesting to note that theoretical studies (e.g. Hachisu et al.\ 2000; Livio 2000; Thoroughgood et al. 2002; see also Justham and Podsiadlowski 2008; Walder et al. 2010) have always looked at U\,Sco and all RNe as likely progenitors of type Ia supernovae. However, observational works on this class of objects seem to show the opposite, i.e. that recurrent novae are not viable SN\,Ia progenitors. On one hand, Della~Valle and Livio (1996) have shown that the frequency of RNe in the Milky Way, M31 and the LMC is significantly smaller (by $\sim$1-2 orders of magnitude) than the supernova Ia rate deduced for these same galaxies. On the other hand, Selvelli et al.\ (2008) have provided observational evidence that T\,Pyx ejects more material than it accretes and therefore it cannot explode as a SN\,Ia. This paper concludes that U\,Sco, hosting a massive ONeMg WD, cannot explode as a SN\,Ia either. Therefore, among RNe, only symbiotic recurrent novae ``survive'' as the possible progenitors of SN\,Ia  (Justham and Podsiadlowski 2008; Di Stefano 2010), explaining at least some of the observed SN\,Ia (Patat et al.\ 2011; Di Stefano 2010). Still, the ultimate fate of RS\,Oph, the prototype object of the symbiotic RNe, remains uncertain (e.g. Osborne et al.\ 2006; Justham and Podsiadlowski 2008).

While the question of how many different stellar systems produce type Ia supernovae  remains unsolved, it has been proven critical to establish not only the mass of the WD and the ejecta, but also the primary star composition in candidate SN\,Ia progenitors.  In the case of RNe, in particular, it will be important to establish whether an ONeMg WD is peculiar to U\,Sco or rather common to the RNe of the same type or to all recurrent novae, as suggested by  Webbink (1990). 

%
%
%

\begin{acknowledgements}
I thank Dr. Bradley Schaefer for having communicated U Sco broad band magnitudes at the time of X-Shooter observations. I also thank Dr. Massimo Della Valle, Dr. Paola Amico and professor Robert E. Williams for kindly revising the paper before submission; and the anonymous referee for the helpful comments and suggestions. 
Finally, I thank the ESO Director General, Prof. Tim de Zeew, for approving the observations under DDT 284.D-5041, which allowed the gathering of the unique X-Shooter spectra. 
\end{acknowledgements}


\begin{thebibliography}{}
\bibitem[2000]{anu} Anupama, G.C., Dewangan G.C., 2000, AJ, 119, 1359
\bibitem[2009]{solarAbu} Asplund M., Grevesse N., Sauval A. J., Scott P., 2009, ARAA, 47, 481 
\bibitem[2010]{brist} Bristow P., Vernet J., et al. 2010, SPIE, 7737E, 34 
\bibitem[1996]{dvl} Della Valle M., Livio M., 1996, ApJ, 473, 240
\bibitem[2002]{nvel} Della Valle M., Pasquini L., Daou D., Williams R.E., 2002, A\&A, 390, 155
\bibitem[2010]{Diaz} Diaz M. P., Williams R. E., Luna G. J., Morales M., Takeda L., AJ, 140, 1860
\bibitem[2010]{Distefano} Di Stefano R., 2010, ApJ, 719, 474
\bibitem[2006]{dodo} D'Odorico S., Dekker H.,  et al. , 2006, SPIE, 6269E, 98
\bibitem[2001]{Dopita} Dopita M.A., Sutherland R.S., 2001, ``Diffuse Matter in the Universe'', ed W. Beiglbock
\bibitem[2000]{Hacisu} Hachisu I., Kato M., Kato T., Matsumoto K., 2000, ApJ, 528, L97
\bibitem[2000]{Hacisu e kato} Hachisu I., Kato M., 2002, AIP Conf. Proceeding, 637, 284
\bibitem[2010]{hel} Helton L.J., Woodward C.E., et al. 2010, AJ, 140, 1347 
\bibitem[2008]{justham} Justham S., Podsiadlowski P., 2008, ASP Conference Series, 401, 161 
\bibitem[1995]{bob95} Kingdon J., Williams R. E., 1997, AJ, 113, 2193
\bibitem[1994]{Livio} Livio M., Truran J.W., 1994, ApJ, 425, 797
\bibitem[2000]{Livio} Livio M., 2000, in {\it Type Ia Supernovae, Theory and Cosmology.}, Cambridge University Press, p.33
\bibitem[2003]{matsumoto} Matsumoto K., Kato T, Hachisu I, 2003, PASJ, 55, 297
\bibitem[2010]{modi} Modigliani A., Goldoni P., et al. , 2010, SPIE, 7737E, 56 
\bibitem[1997]{MeZ} Munari U., Zwitter T., 1997, A\&A, 318, 269
\bibitem[2010]{munari} Munari U., Dallaporta S., Castellani F., 2010, IBVS, 5930
\bibitem[1991]{nomoto1} Nomoto K., Kondo Y., 1991, ApJ, 367, L19
\bibitem[2000]{nomoto} Nomoto K., Umeda H., et al., 2000, AIPC, 522, 35 
\bibitem[2006]{osborne} Osborne J.P., Page K.L., et al. 2006, ApJ, 727, 124 
\bibitem[2006]{ost} Osterbrock D.E., Ferland G.J., 2006, in {\it Astrophysics of gaseous nebulae and active galactic nuclei}, 2$^{nd}$ edition
\bibitem[2011]{patat} Patat F., Chugai N.N., et al., 2011, 530, 63 A\&A
\bibitem[2008]{pod} Podsiadlowski P., 2008, ASP Conf. Series, 401, 63
\bibitem[1995]{Dina} Prialnik D., Livio M., 1995, PASP, 107, 1201 
\bibitem[2010]{schaefer} Schaefer B. E., Pagnotta A., et al. 2010, ApJ, 140, 925
\bibitem[2010]{Schaefer} Schaefer B.E., 2010, arXiv:100
\bibitem[2001]{schwarz02} Schwarz G.J., Shore S.N., et al. 2001, MNRAS, 320, 103 
\bibitem[2002]{schwarz02} Schwarz G.J., 2002, ApJ, 577, 940
\bibitem[2007]{schwarz07} Schwarz G.J., Shore S.N., et al. 2007a, ApJ, 657, 453
\bibitem[2007]{schwarz07} Schwarz G.J., Woodward C.E.,  et al. 2007b, ApJ, 134, 516
\bibitem[2008]{selvelli} Selvelli P., Cassatella A., et al. 2008, A\&A, 492, 787
\bibitem[2003]{novavel} Shore S.N., Bond H. E., et al. 2003, AJ, 125, 1507 
\bibitem[2006]{cn2_s}Shore S.N., 2008, in {\it The Classical Novae}, 2$^{nd}$ edition, Ed. M.F. Bode and A. Evands, (Cambridge Univ. Press), p.203
\bibitem[1992]{starrfield} Starrfield S., Politano M., Truran J. W., Sparks W. M., 1992, in {\it NASA. Goddard Space Flight Center, The Compton Observatory Science Workshop}, p. 377
\bibitem[2001]{Thoroughgood} Thoroughgood T.D., Dhillon V.S., et al. , 2001, MNRAS, 327, 1323
\bibitem[2002]{Thoroughgood} Thoroughgood T.D., Dhillon V.S., et al. 2002 , ASP Conf. Proceeding, 261, 77
\bibitem[1988]{truran} Truran J.W., Livio M., et al., 1988, ApJ, 324, 345
\bibitem[1999]{vanlan} Vanlandingham K.M., Starrfield S., et al., 1999, MNRAS, 308, 577 
\bibitem[2005]{vanlan} Vanlandingham K.M., Schwarz G.J., et al., 2005, ApJ, 624, 914 
\bibitem[2008]{joel} Vernet J., Kerber F., et al. , 2008, SPIE, 7016E, 46
\bibitem[1995]{bible} Warner B., 1995, {\it Cataclysmic Variables Stars}, Cambridge Univ. Press
\bibitem[2009]{wf} Walder R., Folini D. et al., 2010, ASPC, 429, 173 
\bibitem[1990]{webbink} Webbink R. F., 1990, in {\it Physiscs of classical novae}, IAU Colloquium N. 122, LNP, 369, 405
\bibitem[2010]{yamanaka} Yamanaka M., Uemura M., et al.,  2010, PASJ, 62L, 37 
\bibitem[2005]{yaron} Yaron O., Prialnik D., Shara M. M., Kovetz A., 2005, ApJ, 623, 398
\end{thebibliography}
\end{document}